\documentclass[prd,aps,superscriptaddress,showpacs,amssymb,twocolumn]{revtex4}
\usepackage{graphicx,bm}

\def\bacol{\setlength{\arraycolsep}{0pt}}
\def\bec{\begin{center}}
\def\enc{\end{center}}

\newcommand \bew {\begin{widetext}}
\newcommand \enw {\end{widetext}}
\def\ben{\begin{equation}}
\def\ba{\begin{array}}
\def\bea{\r\begin{eqnarray}}
\def\een{\end{equation}}
\def\eea{\end{eqnarray}\er}
\def\ea{\end{array}}
\def\btab{\begin{table}}
\def\btabu{\begin{tabular}}
\def\etab{\end{table}}
\def\etabu{\end{tabular}}
\def\bit{\begin{itemize}}
\def\eit{\end{itemize}}
\def\bef{\begin{figure}[htb]}
\def\befh{\begin{figure}[h!!]}
\def\enf{\end{figure}}

\def\la{\langle}
\def\ra{\rangle}

\def\a{\alpha}
\def\ro{\rho}
\def\k{\kappa}

\def\unit{\mbox{\boldmath 1}}
\def\cA{{\cal A}}
\def\cF{{\cal F}}

\def\cZ{{\cal Z}}

\def\hatU{{\hat{U}}}

\def\hatO{{\hat{\Omega}}}
\def\hatg{{\hat{g}}}
\def\hatom{{\hat{\omega}}}

\def\gb{\beta}

\def\D{\Delta}

\def\e{\epsilon}

\def\om{\omega}

\def\O{\Omega}

\def\s{\sigma}

\def\t{\tau}

\def\Ga{\Gamma}

\def\U{{\cal U}}

\def\shalf{{\scriptstyle{1 \over 2}}}

\def\b1{{\bf 1}}

\def\be{{\bf e}}
\def\b0{{\bf 0}}

\def\bark{{\bar{k}}}
\def\barbbk{{\mbox{\boldmath $\bar{k}$}}}
\def\tv{{\tilde{v}}}

\def\bx{{\mbox{\boldmath $x$}}}

\def\be{{\mbox{\boldmath $e$}}}
\def\bv{{\mbox{\boldmath $v$}}}
\def\bP{{\mbox{\boldmath $P$}}}

\def\bn{{\mbox{\boldmath $n$}}}

\def\sbn{{\mbox{\boldmath $\scriptstyle n$}}}

\def\bl{{\mbox{\boldmath $l$}}}

\def\cos{\hbox{cos}\:}
\def\sin{\hbox{sin}}

\def\nn{\nonumber}
\def\bb{\left(}
\def\eb{\right)}

\def\1{{1}}
\def\mod{\mbox{mod}}
\def\pr{\prime}

\def\U{{\bf U}}

\def\rw{\rule[-5mm]{0mm}{12mm}}
\def\rw0{\rule[0mm]{0mm}{15mm}}
\def\rb{\raisebox{3mm}[0pt]}
\def\rb0{\raisebox{0mm}[0mm][20truemm]{\null}}

\def\r{}
\def\er{}

\newcommand{\cambridge}{DAMTP, CMS, University of Cambridge, Wilberforce Road,
Cambridge CB3 0WA, U.K.}
\newcommand{\edinburgh}{School of Physics, University of
Edinburgh, King's Buildings, Edinburgh EH9 3JZ, U.K.}

\newcommand{\Tr}{\mathop{\mathrm{Tr}}}
\newcommand{\tr}{\mathop{\mathrm{tr}}}
\newcommand{\real}{\mathop{\mathrm{Re}}}

\begin{document}

\title{Perturbation theory vs. simulation for tadpole improvement
  factors in pure gauge theories}

\author{A. \surname{Hart}}
\affiliation{\edinburgh}

\author{R.R. \surname{Horgan}}
\author{L.C. \surname{Storoni}}
\affiliation{\cambridge}

\begin{abstract}
  We calculate the mean link in Landau gauge for Wilson and improved
  $SU(3)$ anisotropic gauge actions, using two loop perturbation
  theory and Monte Carlo simulation employing an accelerated Langevin
  algorithm.  Twisted boundary conditions are employed, with a twist
  in all four lattice directions considerably improving the (Fourier
  accelerated) convergence to an improved lattice Landau gauge. Two
  loop perturbation theory is seen to predict the mean link extremely
  well even into the region of commonly simulated gauge couplings and
  so can be used remove the need for numerical tuning of
  self-consistent tadpole improvement factors. A three loop
  perturbative coefficient is inferred from the simulations and is
  found to be small. We show that finite size effects are small and
  argue likewise for (lattice) Gribov copies and double Dirac sheets.
\end{abstract}

\preprint{Edinburgh 2004/02}


\maketitle

\section{Introduction}
\label{sec_introduction}

Tadpole improvement is now widespread in lattice field theory
\cite{Lepage:1993xa}.
Without it, lattice perturbation theory begins to fail on distance
scales of order 1/20~fm.  Perturbation theory in other
regularisations, however, seems to be phenomenologically successful
down to energy scales of the order of 1~GeV (corresponding to lattice
spacings of 0.6~fm)
\cite{Lepage:1996jw}.

The reason is that the bare lattice coupling is too small
\cite{Lepage:1993xa,Lepage:1996jw}.
To describe quantities dominated by momenta of order the cut-off scale
($\pi/a$), it is appropriate to expand in the running coupling, $\alpha_s$,
evaluated at that scale. The bare coupling, however, deviates markedly
from this and its anomalously small value at finite lattice spacing
can be associated with tadpole corrections
\cite{Lepage:1996jw}.
These tadpole corrections are generally process independent and can
be (largely) removed from all quantities by modifying the action.
This corresponds to a resumming of the perturbative series to yield an
expansion in powers of a new, ``boosted'' coupling that is much closer
to $\alpha_s(\pi/a)$.

Perturbatively this amounts to adding a series of radiative
counterterms to the action. Such a series is obtained by dividing each
gauge link in the action by an appropriate expansion,
$u^{(\mathrm{PT})}$. It is sufficient that this series is known only
up to the loop order of the other quantities we wish to calculate
using the action. 

The factor $u^{(\mathrm{PT})}$ is not unique, but it should clearly be
dominated by ultraviolet fluctuations on the scale of the cut-off. The
two most common definitions are the fourth root of the mean plaquette
(a gauge invariant definition) and the expectation value of the link
in Landau gauge.  Both are successful, although lattice results
suggest some arguments for preferring the latter
\cite{Lepage:1998id}.
In this paper we discuss Landau gauge mean link tadpole improvement. 

For Monte Carlo simulations, each gauge link in the action is divided
by a numerical factor, $u^{(\mathrm{MC})}$. Its value is fixed by a
self--consistency criterion; the value measured in the simulation
should agree with the parameter in the
action. Obtaining such numerical agreement requires computationally
expensive tuning of the action. 
In many cases this cost is prohibitive, such as the zero temperature
tuning of highly anisotropic actions for use in finite temperature
simulations. As non--perturbative phenomena should not affect the
cut-off scale, a sufficiently high order perturbative series should
predict $u^{(\mathrm{MC})}$ such that the subsequent numerical tuning
is unnecessary.

In this paper we present the tadpole improvement factors calculated to
two loop order using lattice perturbation theory. This covers the loop
order of most perturbative calculations using lattice actions. In
addition, we perform Monte Carlo simulations over a range of gauge
couplings extending from the high-$\beta$ regime down to the lattice
spacings used in typical simulations. 
We demonstrate that the two loop formula predicts the numerically
self--consistent $u^{(\mathrm{MC})}$ to within a few digits of the
fourth decimal place and the additional tuning required is minimal
(especially when the action can be rescaled as in
\cite{Drummond:2002yg,Drummond:2003qu}).
For this reason we refer to both $u^{(\mathrm{PT})}$ and
$u^{(\mathrm{MC})}$ as $u$ from hereonin. The small deviations at
physical couplings are shown to be consistent with a third order
correction to $u$ and we infer the coefficient of this. In most
cases no tuning at all is required if the third order is included.

These calculations are carried out for two $SU(3)$ lattice gauge
actions; the Wilson action and a first order Symanzik improved action.
Isotropic and anisotropic lattices are studied and interpolations of
the coefficients with the anisotropy are given.

The structure of this paper is as follows. We first discuss the
perturbative calculation in Section~\ref{sec_pert_th}. In
Section~\ref{sec_mc} we describe high-$\beta$ Monte Carlo simulations,
and use the results to obtain higher order coefficients in the
perturbative expansion, before concluding in Section~\ref{sec_conc}. A
brief description of twisted boundary conditions is given in
Appendix~\ref{app_twbc}.

The results presented here extend and, in some cases, correct the
preliminary results presented in
\cite{Drummond:2002kp}.
Extension of this work to actions including fermions is being carried
out and will be reported in a future publication.

\section{Perturbation theory}
\label{sec_pert_th}

We write the Landau gauge mean links
\begin{equation}
u_\mu = \left\langle \tr U_\mu \right\rangle
\end{equation}
(where $N$ is the number of colours and $\tr \equiv \frac{1}{N} \Tr$) as
\begin{equation}
u_\mu = 1 + a^{(1)}_\mu g^2 + a^{(2)}_\mu g^4 + {\cal O}(g^6) \; .
\end{equation}
Using an anti-Hermitian gauge potential, $U_\mu = \exp g A_\mu$, the
mean Landau link in perturbation theory is
\begin{equation}
\left\langle \tr U_\mu \right\rangle =
1 + g^2 \left\langle \tr \frac{A_\mu^2}{2} \right\rangle +
g^4 \left\langle \tr \frac{A_\mu^4}{24} \right\rangle
+ {\cal O}(g^6) \; .
\end{equation}
We use twisted boundary conditions to regulate the gluon zero mode in
a gauge invariant manner.  A brief \textit{resum\'{e}} of relevant
results and definitions is given in Appendix~\ref{app_twbc} and the
reader is referred to
\cite{luwe,Drummond:2002yg}
for a fuller discussion.

\begin{figure}[t]
\bec
\includegraphics[width=50mm,clip]{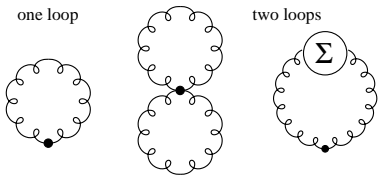}
\enc
\caption{\label{fig_meanlink_diags}
Feynman diagrams for the mean link.
}
\end{figure}

\begin{figure}[t]
\bec
\includegraphics[height=50mm,clip]{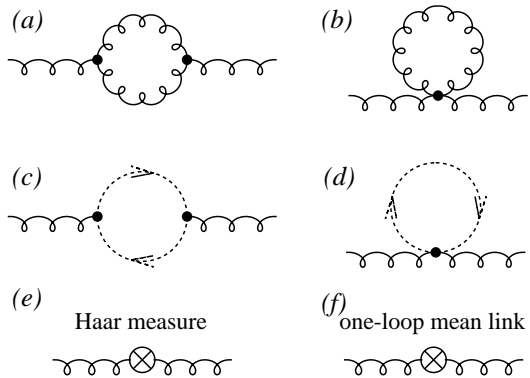}
\enc
\caption{\label{fig_selfenergy_diags}
Feynman diagrams for the gluon self energy, $\Sigma_{\mu \nu}$. 
Feynman rules are discussed in \cite{Drummond:2002yg}.
}
\end{figure}

\subsection{one loop}

The one loop contribution comes from
\begin{eqnarray}
\left\langle \tr A_\mu^2 \right\rangle
& = & \frac{1}{N} \frac{1}{V_{TW}^2} \sum_{\bm{k},\bm{p}}
e^{i \left(\bm{k} + \bm{p}\right) \cdot \bm{x}}
\nonumber \\
&& \times 
\mathop{\mathrm{Tr}} \left[ \Gamma(\bm{k}) \Gamma(\bm{p}) \right]
\tilde{A}_\mu(\bm{k}) \tilde{A}_\mu(\bm{p}) \; .
\end{eqnarray}
The one loop coefficient is thus the sum over the tree level
propagator in Landau gauge (see Eqn.~(18) of
\cite{Drummond:2002yg}),
\begin{equation}
a^{(1)}_\mu = \frac{1}{2 V_{TW}} \sum_{\bm{k}} 
z^{\frac{1}{2}(\bm{k},\bm{k})}
G_{\mu\mu}(\bm{k}) \; ,
\label{eqn_one_loop}
\end{equation}
shown in Fig.~\ref{fig_meanlink_diags}. We note here that the phase in
the integrand is precisely that of the inverse propagator in the
twisted formalism (see Eqn.~(63) of
\cite{Drummond:2002yg}),
so $a^{(1)}_\mu$ is real.

\begin{table}[t]

\caption{\label{tab_oneloop} One loop coefficients for
various anisotropies, $\chi$, extrapolated to infinite lattice size,
$L$.}

\begin{center}
\begin{ruledtabular}
\begin{tabular}{cccc}
action &
anisotropy &
$a^{(1)}_s$ &
$a^{(1)}_t$ \\
\hline
Wilson &
1.0 & -0.077467 & -0.077467 \\
 &
1.25 & -0.087488 & -0.051452 \\
 &
1.5 & -0.095096 & -0.036299 \\
 &
2.0 & -0.105351 & -0.020351 \\
 &
2.5 & -0.111543 & -0.012886 \\
 &
3.0 & -0.115452 & -0.008851 \\
 &
4.0 & -0.119871 & -0.004899 \\
 &
6.0 & -0.123393 & -0.002142 \\
 &
8.0 & -0.124713 & -0.001197 \\
\hline
Symanzik &
1.0 & -0.063049 & -0.064664 \\
improved &
1.25 & -0.071472 & -0.042877 \\
 &
1.5 & -0.077940 & -0.030150 \\
 &
2.0 & -0.086772 & -0.016934 \\
 &
2.42 & -0.091467 & -0.011462 \\
 &
2.5 & -0.092171 & -0.010720 \\
 &
2.78 & -0.094281 & -0.008616 \\
 &
3.0 & -0.095620 & -0.007365 \\
 &
3.5 & -0.097922 & -0.005363 \\
 &
4.0 & -0.099521 & -0.004077 \\
 &
5.0 & -0.10152 & -0.002584 \\
 &
6.0 & -0.10266 & -0.001784 \\
 &
8.0 & -0.10384 & -0.000997 \\
\end{tabular}
\end{ruledtabular}
\end{center}

\end{table}

\subsection{two loops}

The two loop contribution arises from two sources. The first is from
the above integral, Eqn.~(\ref{eqn_one_loop}), where the tree level
propagator is replaced by the one loop self energy bubble shown in
Fig.~\ref{fig_selfenergy_diags}, giving a term
\bea
a_\mu^{(2)} & = &
\frac{1}{2 V_{TW}} \sum_{\bm{k}} 
z^{\frac{1}{2}(\bm{k},\bm{k})}
G_{\mu \alpha}(\bm{k}) \Sigma_{\alpha \beta}(\bm{k})
G_{\beta \mu}(\bm{k}) 
\nn \\
&& + \mathrm{~Eqn.~(\ref{eqn_fig8_terms}).}
\label{eqn_vegas_terms}
\eea
As $\Sigma_{\alpha \beta}$ has the phase of the inverse propagator,
this expression is also explicitly real.

\begin{table*}

\caption{\label{tab_twoloops}Two loop coefficients for various
anisotropies, $\chi$, extrapolated to infinite lattice size, $L$.}

\begin{center}
\begin{ruledtabular}
\begin{tabular}{cccccc}
action &
anisotropy &
$a^{(2)}_s$ &
$a^{(2)}_t$ &
$b^{(2)}_s$ &
$b^{(2)}_t$ \\
\hline
Wilson &
1.0 & -0.021093 (63) & -0.021093 (63) & 0.002912 (63) & 0.002912 (63) \\
 &
1.25 & -0.023891 (74) & -0.013222 (91) & 0.003573 (74) & 0.002930 (91) \\
 &
1.5 & -0.025721 (95) & -0.008913 (63) & 0.004861 (95) & 0.002760 (63) \\
 &
2.0 & -0.027669 (91) & -0.004730 (37) & 0.007771 (91) & 0.002116 (37) \\
 &
2.5 & -0.02827 (11) & -0.002900 (23) & 0.01049 (11) & 0.001578 (23) \\
 &
3.0 & -0.02839 (12)  & -0.001916 (16) & 0.01262 (12)  & 0.001228 (16) \\
 &
4.0 & -0.02811 (13)  & -0.001008 (14) & 0.01558 (13)  & 0.000778 (14) \\
 &
6.0 & -0.02718 (27)  & -0.0004080 (42) & 0.01876 (27) & 0.0000377 (42) \\
\hline
Symanzik &
1.0 & -0.01327 (23) & -0.01435 (24) & 0.00273 (23) & 0.00206 (24) \\
improved &
2.0 & -0.01831 (34) & -0.003444 (63) & 0.00574 (34) & 0.001251 (63) \\
 &
3.0 & -0.01931 (50) & -0.001456 (30) & 0.00882  (50) & 0.000711 (30) \\
 &
4.0 & -0.02011 (53) & -0.0010482 (181) & 0.01001 (53) & 0.0001857 (181) \\
\end{tabular}
\end{ruledtabular}
\end{center}

\end{table*}

The second term arises from the tree level portion of the $A_\mu^4$
term:
\begin{eqnarray}
\left\langle \tr A_\mu^4 \right\rangle & = &
\frac{1}{N} \frac{1}{V_{TW}^4} \sum_{\bm{k}_1...\bm{k}_4}
\exp \left[ i (\bm{k}_1 + ...+ \bm{k}_4 ) \cdot \bm{x}
\right]
\nonumber \\
&& \times \mathop{\mathrm{Tr}} \left[ \Gamma(\bm{k}_1) ... 
\Gamma(\bm{k}_4) 
\right]
\tilde{A}_\mu (\bm{k}_1) ... \tilde{A}_\mu(\bm{k}_4) .
\end{eqnarray}
Using Eqn.~(\ref{eqn_tw_alg}), this reduces to
\bea
a_\mu^{(2)} & = & \mathrm{~Eqn.~(\ref{eqn_vegas_terms})}
\nn \\
&& + ~
\frac{1}{24 V_{TW}^2} \sum_{\bm{k},\bm{p}} 
\left( 2 + z^{\langle\bm{k},\bm{p}\rangle} \right)
z^{\frac{1}{2}(\bm{k},\bm{k})} G_{\mu\mu}(\bm{k}) 
\nn \\
&& ~~~ \times
z^{\frac{1}{2}(\bm{p},\bm{p})} G_{\mu\mu}(\bm{p}) \; .
\label{eqn_fig8_terms}
\eea
Since the integrand is symmetric in $\bm{k}$ and $\bm{p}$,
only the real, cosine part of
$z^{\langle\bm{k},\bm{p}\rangle}$ survives and the integral is
real.

As $\langle\bm{k},\bm{p}\rangle$ depends only on the twist
components of $\bm{k},\bm{p}$, we note that if
\begin{equation}
\sum_{\bar{\bm{k}}} z^{\frac{1}{2}(\bm{k},\bm{k})} 
G_{\mu\mu}(\bm{k})
\end{equation}
is stored for each possible value of the associated twist vector (see
Appendix for the definition of $\bar{\bm{k}}$), the second expression
in Eqn.~(\ref{eqn_fig8_terms}) can be reduced to the product of two
one--loop calculations.

\begin{table*}

\caption{\label{tab_contribs} Contributions to the two loop
coefficients, $a^{(2)}_{s,t}$. The notation used
in labelling the contributions is explained in the text.}

\begin{center}
\begin{ruledtabular}
\begin{tabular}{ccccccc}
action &
coefficient &
anisotropy &
VEGAS &
Figure of 8 &
$a^{(1)}_s \Delta S_{ct}$ &
$\Delta S_{ct}$ \\
\hline
Wilson & $a^{(2)}_s$ &
1.0 & -0.022969 (63) & 0.001875 \\
 & &
1.25 & -0.026283 (74) & 0.002392 \\
 & &
1.5 & -0.028547 (95) & 0.002826 \\
 & &
2.0 & -0.031138 (91) & 0.003468 \\
 & &
2.5 & -0.03216 (11) & 0.003888 \\
 & &
3.0 & -0.03255 (12) & 0.004166 \\
 & &
4.0 & -0.03260 (13) & 0.04488 \\
 & &
6.0 & -0.03194 (27) & 0.004759 \\
\cline{2-5}
& $a^{(2)}_t$ &
1.0 & -0.022969 (63) & 0.001875 \\
 & &
1.25 & -0.014049 (91) & 0.000827 \\
 & &
1.5 & -0.009323 (63) & 0.000410 \\
 & &
2.0 & -0.004859 (37) & 0.000130 \\
 & &
2.5 & -0.002952 (23) & 0.000052 \\
 & &
3.0 & -0.001941 (16) & 0.000025 \\
 & &
4.0 & -0.001015 (14) & 0.000008 \\
 & &
6.0 & -0.0004095 (42) & 0.000002 \\
\hline
Symanzik & $a^{(2)}_s$ &
1.0 & -0.01232 (23) & 0.0012422 & -0.0021920 & 0.034766 \\
improved & &
2.0 & -0.01693 (34) & 0.0023529 & -0.0037334 & 0.043026 \\
 & &
3.0 & -0.01780 (50) & 0.0028573 & -0.0043696 & 0.045697 \\
 & &
4.0 & -0.01855 (53) & 0.0030951 & -0.0046588 & 0.046811 \\
\cline{2-7}
& $a^{(2)}_t$ &
1.0 & -0.01468 (24) & 0.00130671 & -0.00097386 & 0.015446 \\
 & &
2.0 & -0.003193 (63) & 0.00008961 & -0.00034017 & 0.0039203 \\
 & &
3.0 & -0.00131 (30) & 0.00001695 & -0.00016325 & 0.0017073 \\
 & &
4.0 & -0.0009590 (181) & 0.00000520 & -0.00009436 & 0.0009481 \\
\end{tabular}
\end{ruledtabular}
\end{center}

\end{table*}
\subsection{Numerical integration}

We consider the lattice Wilson action (W) and the Symanzik improved action
(SI) defined in
\cite{alea}, 
both with tadpole improvement in the spatial ($s$) and temporal ($t$)
directions.
{\bacol
\bea
S_W && (\beta_0,\chi_0,u_s,u_t) =
- \frac{\beta_0}{\chi_0}
\sum_{x,s>s'} 
\frac{P_{s,s'}}{u_s^4} -
\beta_0 \chi_0
\sum_{x,s} \frac{P_{s,t}}{u_s^2u_t^2}
\nonumber \\
S_{SI} && (\beta_0,\chi_0,u_s,u_t) =
\nonumber \\
&& - \frac{\beta_0}{\chi_0} \sum_{x,s>s'} 
{ \left( \frac{5}{3}\frac{P_{s,s'}}{u_s^4} - 
\frac{1}{12}\frac{R_{ss,s'}}
{u_s^6}-\frac{1}{12}\frac{R_{s's',s}}{u_s^6}\right)
}
\nonumber \\
&&  - \beta_0 \chi_0 \sum_{x,s}
{ \left( \frac{4}{3}\frac{P_{s,t}}
{u_s^2u_t^2}-\frac{1}{12}\frac{R_{ss,t}}{u_s^4u_t^2}\right)
} \; ,\label{actions}
\eea
}
where $s,s'$ run over spatial links; $P_{s,s'}$ and $P_{s,t}$ are
$1\times 1$ plaquettes; $R_{ss,s'}$ and $R_{ss,t}$ are $2\times 1$
loops; and $\chi_0$ is the bare anisotropy.

It is clearly inconvenient to work perturbatively with actions whose
parameters are implicit functions of the expansion parameter
$g_0^2$. To circumvent this problem we define
\begin{equation}
\beta = \beta_0/(u_s^3 u_t) \; , ~~g^2 = g_0^2 u_s^3 u_t \; ,
~~\chi = \chi_0 u_s/u_t \; ,\label{sc_couplings}
\end{equation}
and find
{\bacol
\bea
S_W && (\beta_0,\chi_0,u_s,u_t) = S_W(\beta,\chi,u_s=u_t=1)\;,
\nonumber \\
S_{SI} && (\beta_0,\chi_0,u_s,u_t)~=~S_{SI}(\beta,\chi,u_s=u_t=1)~+
\nn \\
&& g^2 \Delta S_{SI}  + {\cal O}(g^4) \; ,
\nonumber \\
\Delta S_{SI} && = d_s \Delta S_{ct} 
\nonumber \\
&& = d_s \beta \sum_{\stackrel{x}{s>s'}}
\frac{1}{6} \left( \frac{\left[ R_{ss,s'}+R_{s's',s} \right]}{\chi} +
\chi R_{ss,t}\right),
\label{sc_actions}
\eea
}
where $d_s$ is the one loop coefficient in the (self-consistent)
perturbative expansion of the tadpole improvement factor. For mean
Landau link improvement $d_s = a^{(1)}_s$.

Perturbative expansions will initially be calculated as power series
in $g^2$ and the expansion then re-expressed as a series in $g_0^2$.
 
The Feynman rules are obtained by perturbatively expanding the lattice
actions using an automated computer code written in {\sc Python}
\cite{luwe,Drummond:2002kp}.
The additional vertices associated with the ghost fields and the
measure for Landau gauge are given in
\cite{luwe,Drummond:2002yg}.
The Feynman diagrams for the one and two loop calculations are shown
in Figs.~\ref{fig_meanlink_diags} and~\ref{fig_selfenergy_diags}.

The one loop integration is carried out by direct summation of all
twisted momentum modes for hypercubic lattices with $4 \le L_\mu \le
32$. To speed up the approach to infinite volume, the momenta are
``squashed'' in the directions with periodic boundary conditions using
the change of variables $\bm{k} \rightarrow \bm{k}^\prime$ suggested by
L\"{u}scher and Weisz
\cite{luwe}
\ben
k^\prime_\mu = k_\mu - \a_\mu \; \sin(k_\mu)\;, \label{cofv}
\een
giving an integrand with much broader peaks which is easier to
evaluate numerically. It is easy to see that a reasonable choice of
parameter is $\a_\mu \sim 1-(\chi L_\mu)^{-1}$ and significantly
reduced the dependence on $L$. The calculations were all possible on a
workstation.

All results were extrapolated to infinite volume using a fit function
of the form $c_0 + c_1/L^2$, which worked extremely well for $L \ge
16$. As a further check, the coefficients in the expansion of the
spatial mean link should extrapolate to the same value in twisted and
periodic directions. We found this to be the case, with the
discrepancy much smaller than the number of significant figures quoted
in the results in this paper.

We calculate the two loop coefficient in three parts.  The ``figure of
8'' diagram from Eqn.~(\ref{eqn_fig8_terms}) is calculated by mode
summation using the reduction described above.

The remaining two loop calculations arising from
Fig.~\ref{fig_selfenergy_diags}$(a-d)$ (and from the trivial
Fig.~\ref{fig_selfenergy_diags}$e$) were carried out using {\sc
Vegas}, a Monte Carlo estimation program
\cite{lepage78,numrec}.
With a version of {\sc Vegas} parallelised using the MPI protocol,
these computations were carried out on between 64 and 256 processors
of a Hitachi SR2201 supercomputer. Each run took of the order of 24
hours.  We label the results from this calculation ``VEGAS''. We used
the 2--twist boundary conditions described in the Appendix. The
lattice was of infinite extent in the untwisted $\mu=3,4$ directions,
and of size $L \in \{4,8,16,32 \}$ in the twisted $\mu=1,2$
directions.  Again, the L\"uscher-Weisz squashing was applied to the
untwisted momentum components.

Finally, for the Symanzik improved actions there is a counterterm
insertion, Fig.~\ref{fig_selfenergy_diags}$f$, arising from $\Delta
S_{ct}$, which requires a one loop calculation carried out by mode
summation. Multiplying this result by the appropriate $d_s$ gives the
full contribution of the counterterm.

We show the one loop results in Table~\ref{tab_oneloop} and the two
loop results for $d_s = a^{(1)}_s$ in Table~\ref{tab_twoloops}. As it
is possible that results for the mean link in Landau gauge from an
action with a different tadpole improvement scheme might be desired
(or an action with no improvement at all, when $d_s=0$), the
contributions of the various parts of the two loop calculation are
shown in Table~\ref{tab_contribs} to allow the reader to construct the
appropriate two loop contribution (the one loop numbers being
unchanged)
\cite{Drummond:2003qu}.

\begin{table*}

\caption{\label{tab_expansions}\small
Fits to perturbative expansion coefficients as functions of $\chi$, as
defined in the text.}

\begin{center}
\begin{ruledtabular}
\begin{tabular}{cccccc}
action & quantity &
const. &
$1 / \chi$ &
$1 / \chi^2$ &
$\log \chi / \chi$ \\
\hline
Wilson & $a_s^{(1)} = b_s^{(1)}$ & -0.120648 & 0.103142 &
-0.0599588 & -0.0614259 \\
& $a_t^{(1)} = b_t^{(1)}$ & 0 & -0.0168915 & -0.0609128 & 0.00881346 \\
\cline{2-6}
 & $a_s^{(2)}$ & -0.0174148 & 0.0198738 & -0.0235448 & -0.0410654 \\
 & $a_t^{(2)}$ & 0 & 0.00384188 & -0.024975 &-0.00118467 \\
 & $b_s^{(2)}$ & 0.0296102 & -0.041532 & 0.0148529 & -0.0134185 \\
 & $b_t^{(2)}$ & 0 & 0.0180304 & -0.0152916 & -0.00847315 \\
\hline
Symanzik & $a_s^{(1)} = b_s^{(1)}$ & -0.10043 & 0.0915721 & -0.0541941 &
-0.0536973 \\
improved & $a_t^{(1)} = b_t^{(1)}$ & 0 & -0.012287 & -0.0526966 &
0.00656986 \\
\cline{2-6}
& $a_s^{(2)}$ & -0.0300447 & -0.0215242 & 0.038299 & 0.0372853 \\
& $a_t^{(2)}$ & 0 & 0.00708635 & -0.0214381 & -0.00419861 \\
& $b_s^{(2)}$ & 0.00122385 & -0.0728384 & 0.0743446 & 0.0644862 \\
& $b_t^{(2)}$ & 0 & 0.0129186 & -0.0108602 & -0.00676255 \\
\end{tabular}
\end{ruledtabular}
\end{center}

\end{table*}

\begin{figure}[b]
\bec
\includegraphics[height=70mm,clip]{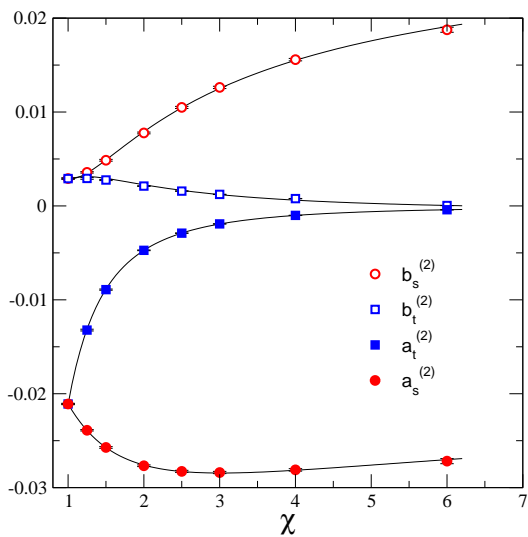}
\enc
\caption{\label{fig_wil_twoloop}
Two loop coefficients to the Landau mean link for the Wilson action.
}
\end{figure}

\begin{figure}[b]
\bec
\includegraphics[height=70mm,clip]{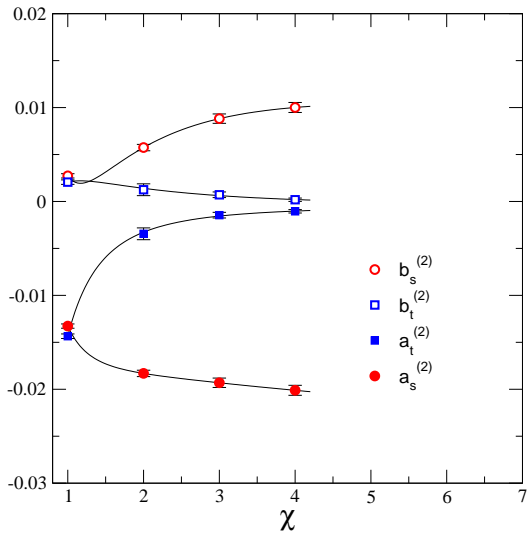}
\enc
\caption{\label{fig_si_twoloop}
Two loop coefficients to the Landau mean link for the 
tadpole improved Symanzik action.
}
\end{figure}

\subsection{Resumming the series}

Perturbatively speaking, tadpole improvement resums the perturbative
expansion in the hope of reducing truncation errors in the
series. This amounts to using $g_0^2$ as the expansion parameter,
rather than $g^2$. Writing 
{\bacol
\begin{equation}
u_l  = \left\{ 
\begin{array}{lll}
1 + a^{(1)}_l g^2     & + a^{(2)}_l g^4     & + {\cal O}(g^6) \; , \\ 
1 + b^{(1)}_l g_0^2 & + b^{(2)}_l g_0^4 & + {\cal O}(g_0^6) \; ,
\end{array}
\right.
\end{equation}
}
we find 
\begin{eqnarray}
b^{(1)}_l & = & a^{(1)}_l \; ,
\nonumber \\
b^{(2)}_l & = & a^{(2)}_l + a^{(1)}_l
(3a^{(1)}_s+a^{(1)}_t) \; .
\label{eqn_boost}
\end{eqnarray}
The coefficients $b^{(2)}_{s,t}$ for mean Landau link tadpole
improvement are given in Table~\ref{tab_twoloops}.

Finally, we carry out interpolating fits in the anisotropy,
$\chi$, to all perturbative coefficients using the function
\begin{equation}
c_0 + \frac{c_1}{\chi} + \frac{c_2}{\chi^2} + c_3 \frac{\log \chi}{\chi} \; .
\label{ufit}
\end{equation}
These fits work well. The fit parameters are given in
Table~\ref{tab_contribs} and are shown in Figs.~\ref{fig_wil_twoloop}
and~\ref{fig_si_twoloop}. It should be noted when using these results
that $\chi$ is the anisotropy {\em after} the majority of tadpole
factors have been scaled out of the action, as is shown in Eqn. (\ref{sc_actions}).

\section{Monte Carlo simulation}
\label{sec_mc}

By comparing the truncated two loop perturbative series obtained above
with results from high-$\beta$ Monte Carlo simulations, higher order
coefficients in the perturbative expansion may be inferred
\cite{Trottier:2001vj}.
\subsection{Implementing the Langevin update}

Field configurations were generated using a 2nd order Runge-Kutta
Langevin updating routine. The implementation of the Langevin
evolution is such that any pure gauge action can be simulated by
simply specifying the list of paths and the associated couplings. The
group derivative for each loop is then computed automatically by an
iterative procedure which moves around the loop link by link
constructing the appropriate traceless anti-Hermitian contribution to
the Langevin velocity. This is the most efficient implementation,
minimising the number of group multiplications needed and can be
applied whenever the quantity to be differentiated is specified as a
(closed) Wilson path.

The method is as follows. We specify a given link by a start site,
$\bx$, and a signed direction, $\mu$:
\bea
U_\mu(\bx) & ~~~~ & \mu > 0 \; ,
\nn \\
U_{-\mu}^\dagger(\bx + \be_\mu) && \mu < 0 \; .
\eea
The $\be_\mu$ for $\mu > 0$ are vectors of the unit cell and
$\be_{\mu} = - \be_{-\mu}$ for negative $\mu$.

The Wilson path, $W(\bx,U)$, of length $M$ is a unitary matrix
specified by a starting point $\bx$ (where it is initially stored) and
an ordered list of $M$ links in directions $( \mu_1, ... ,
\mu_M)$. Explicitly,
\ben
W(\bx,U) = \prod_{i \in I_M} \; 
U_{\mu_i}(\bx + \bl_{i-1}) \; , 
\een
with $I_M = (1,2,\ldots,M)$ and
\ben
\bl_i =
\left\{
\begin{array}{l@{~~~~}l}
0 \; , & i = 0 \; , \\
\displaystyle \sum_{j=1}^{i} \be_{\mu_j} \; , & 0 < i \le M-1 \; ,
\end{array}
\right.
\een
The contribution to the pure glue action from closed loops is then
\ben
S(W) = g_W \sum_{\bm{x}} \real \Tr(W(\bx,U)) \; ,
\een
where $g_W$ is the associated coupling constant and $W(\bx,U)$
contributes to the Langevin velocity $\bv(\bx_j,U)$ at all sites
$\bx_j=\bx+\bl_j$ for $j=0\ldots M-1$
\ben
\bv(\bx_j,U) = g_W \cA [W_j(\bx_j,U)] \; ,
\een
where the operator $\cA$ projects onto the $SU(N)$ algebra of
anti-hermitian, traceless matrices. If $I^{(j)}_M$ is the $j$-th left
cyclic permutation of $I_M$,
\ben
W_j(\bx_j,U) = \prod_{i \in I^{(j)}_M} \; U_{\mu_i}(\bx+\bl_{i-1}) \; .
\een
Clearly, $W_j(\bx_j,U)$ can be obtained from $W_{j-1}(\bx_{j-1},U)$
stored at site $\bx_{j-1}$ by only two matrix multiplies and then the
result is stored at site $\bx_j$. In principle, the whole lattice can
be treated in parallel. The outcome is that the number of matrix
multiplications required to compute $\bv$ is $3(M-1)$ rather than the
$M(M-1)$ for \textit{na\"{\i}ve} construction. This improvement is
particularly relevant in application to improved QCD actions with
dynamical fermions.

The 2nd order Runge-Kutta algorithm used is a two step update:
%
\bea
\hatU^{(1)}_\mu(\t,\bx) & = & R^{(1)}(\t,d\t,\bx)\,\hatU_\mu(\t,\bx)
\; , \nn \\
\hatU_\mu(\t+d\t,\bx) & = &R^{(2)}(\t,d\t,\bx)\,\hatU_\mu(\t,\bx)
\; , \nn \\
R^{(1)}(\t,d\t,\bx) & = & \exp(u^{(1)}(\t,d\t,\bx))
\; , \nn \\
R^{(2)}(\t,d\t,\bx) & = & \exp(u^{(2)}(\t,d\t,\bx))
\; , \nn \\
u^{(1)}(\t,d\t,\bx) & = & d\t/2\,\bv(\bx,\hatU(\t)) + 
\nn \\
& & \sqrt{d\t}\,\eta^{(1)}(\t,\bx)
\; , \nn \\
u^{(2)}(\t,d\t,\bx) & = & 
d\t\,C_+\,\bv(\bx,\hatU^{(1)}(\t)) + 
\nn \\
& & \sqrt{d\t\,C_-} \; (\eta^{(1)}(\t,\bx)+\eta^{(2)}(\t,\bx)) \; .
\nn \\
\eea
%
Here, $C_{\pm} = 1 \pm C_N/12$, with $C_N$ being the adjoint Casimir
for $SU(N)$ ($C_3 = 3$). The
$\eta^{(p)}(\t,\bx)$ for $p=1,2$ are $N\times N$
anti-hermitian Gaussian random matrices:
\ben
\eta^{(p)}(\t,\bx) = \eta^{(p)}_a(\t,\bx) T_a \; ,
\een
[with $T_a$ being the (anti-hermitian) generators of $SU(N)$]. They
satisfy
\ben
\left\la\;\eta^{(p)}_a(\t,\bx)\,\eta^{(p^\pr)}_{a^\pr}(\t^\pr,\bx^\pr)
\; \right\ra = 
\delta_{p p^\pr}\,\delta_{\t \t^\pr}\,\delta_{\bm{x}
\bm{x}^\pr}\,\delta_{\a \a^\pr} \; .
\een 
\subsection{Twisted BC and tunnelling}

All 2--, 3-- and~4--twist boundary conditions were implemented for
simulation.  The manner of the implementation is as described in
\cite{luwe}.
An explicit representation of the $\O$ matrices is not needed for the
update for the pure gauge (quenched) update since the field simulated
is $\hatU$ which is related to the perturbative field $\U$ by the
redefinition
\bea
\hatU_\mu(\bx) & = & U_\mu(\bx)\hatO_\mu(\bx)
\nn \\
\hatO_\mu(\bx) & = & \left\{
\ba{lcl}\unit &~~~~~~& x_\mu \ne L_\mu \; ,
\\
\O_\mu&&x_\mu = L_\mu \; .
\ea 
\right.
\eea
The action in terms of the $\{\hatU\}$ is identical to the untwisted
action except that a loop whose projection onto the $(\mu,\nu)$-plane
($\mu<\nu$) encircles the point $(L_\mu+1/2,L_\nu+1/2)$ has an
additional factor of $(z_{\mu \nu})^{-c}$ where $c$ is the integer
winding number [\textit{i.e.} the sense of the circulation: $c > 0$
$(c < 0)$ for anti-clockwise (clockwise) circulation of $|c|$
turns]. This weights the corresponding contribution to the action by
the factor $\cos (2\pi cn_{\mu\nu}/N)$ compared with the untwisted
case. An explicit representation of $\O$ would be needed for a
simulation with dynamical fermions (and, as we shall see, for gauge
fixing).

\begin{figure}[b]
\includegraphics[height=60mm,clip]{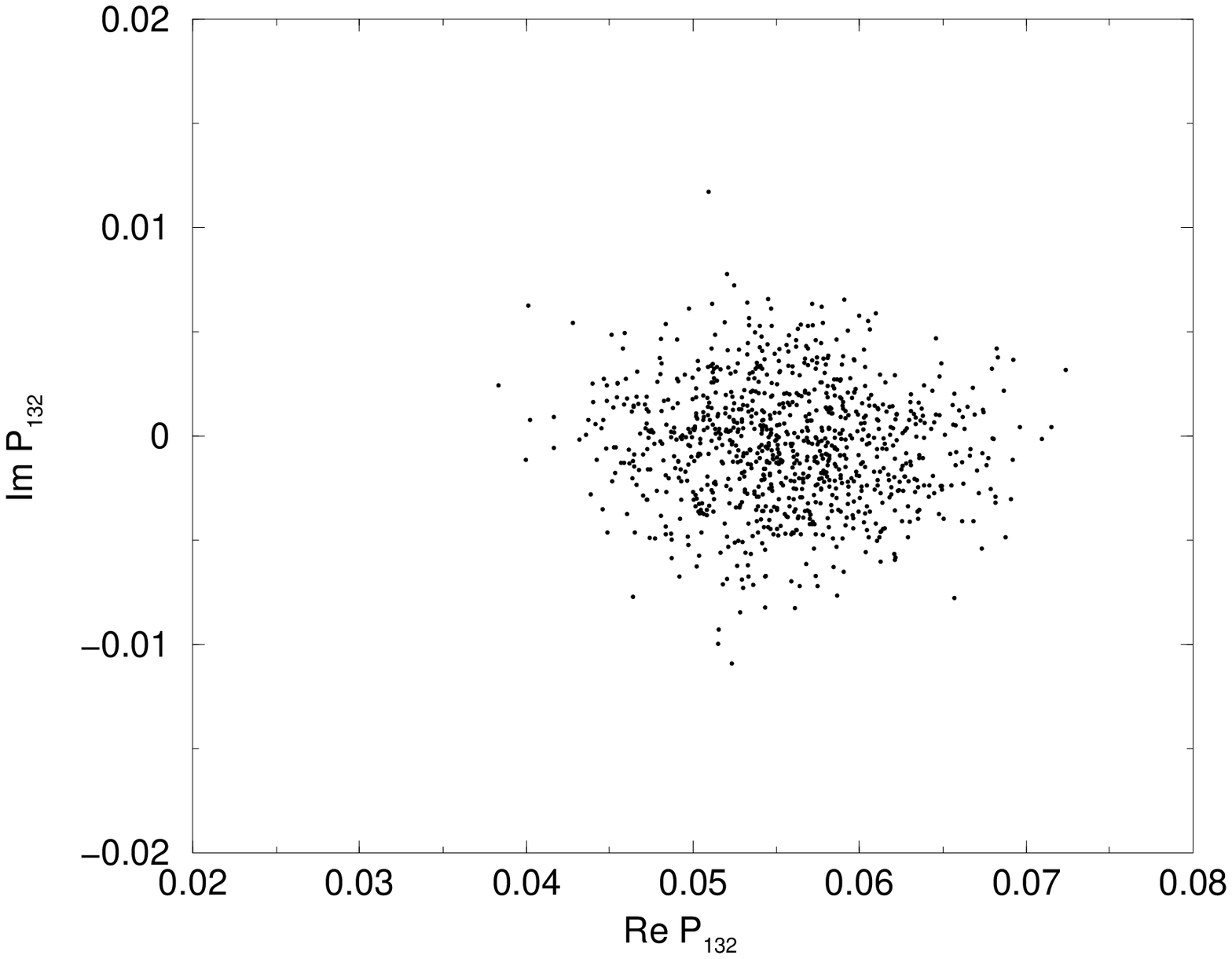}

\includegraphics[height=60mm,clip]{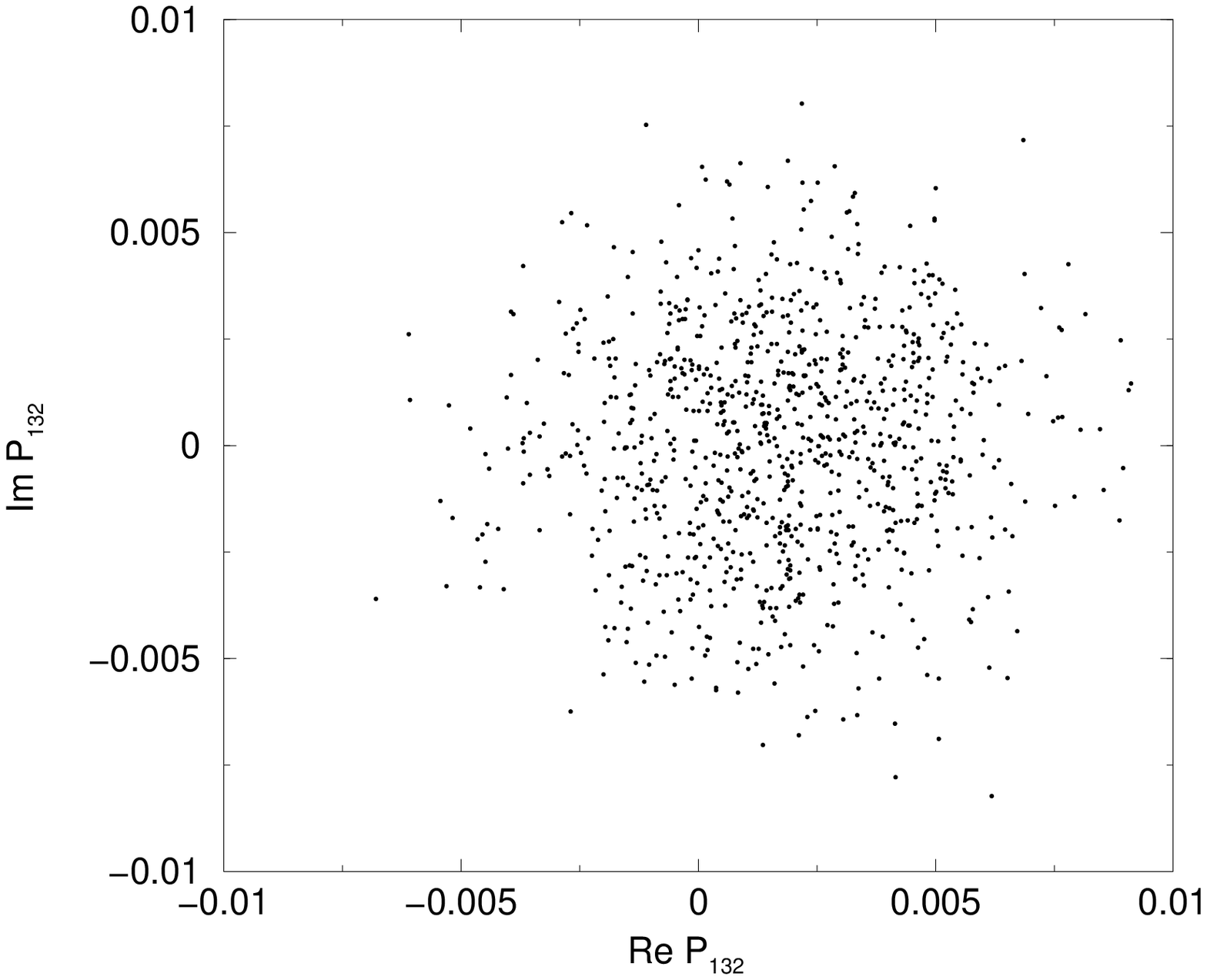}
\caption{\label{fig_p132} The scatter plot for the expectation for
Polyakov line $P_{132}$ for $\beta=10$ (top plot) and $\beta=6$
(bottom plot). Although the mean value is becoming small as $\beta$
decreases, both distributions are centred on a point on the real axis
and show no sign of tunnelling to centre-equivalent distributions.  }
\end{figure}

For twisted boundary conditions the number of different $Z(N)$
phases is $N^2$ 
\cite{gonzalez97} 
and these are distinguished by order parameters which are the set of
independent non-zero Polyakov lines winding around the lattice. For
example, one such Polyakov line in the 4--twist case is
$P_{132}(\bx)$ where
\bea
\bP_i(\bx,\hatU) & = & \prod_{n_i=0}^{L_i}\hatU_i(\bx+n_i\be_i) \; , 
\nn\\
P_{ijk}(\bx,\hatU) & = & \Tr [ 
\bP_i(\bx,\hatU) \bP_j(\bx,\hatU) \bP_k(\bx,\hatU) ] \; . 
\label{p132}
\eea 
Here $i,j,k$ are signed directions as above.  Tunnelling can occur
between different $Z_N$ vacua, corresponding to the multiplication of
all links on a space-like slice by an element $z$ from the centre
group of $SU(N)$.

Twisted boundary conditions create a barrier between the different
$Z(N)$ phases and the zero mode, responsible for non-perturbative
effects
\cite{lusc:1983},
is absent. This is a prerequisite for comparison with perturbation
theory and the measurement of coefficients of terms in the series of
higher order than those calculated.
The study in reference 
\cite{Trottier:2001vj}
shows that with 3--twist boundary conditions tunnelling is absent for
$\beta \ge 9$, even on small lattices of $4^4$. The probability for
tunnelling then reduces as the lattice volume increases. We use a $8^3
\times 16$ lattice and with 4--twist boundary conditions we expect
tunnelling to be rare or absent for lower values of $\beta$.  We can
obtain an idea of whether it is occurring by plotting the
temporal history of a non-zero Polyakov line such as $P_{132}$. The
results are shown in Fig.~\ref{fig_p132} for $\beta=9.0$ and
$\beta=6.0$.  The density of points in both cases is centred on a
point on the real axis and no points are found in the
centre-equivalent positions on the lines in directions
$(-1/2,\pm\sqrt{3}/2)$.  We conclude that tunnelling events are absent
and that we can be confident that the influence of non-perturbative
effects on our measurements has been minimised. For this reason, and
on account of the convergence of the gauge fixing discussed below, we
restricted our simulations to the 4--twist boundary conditions only.

\subsection{Fixing to Landau gauge}

To fix each configuration to Landau gauge we maximise the
corresponding gauge function with respect to gauge transformations. In
the continuum this function is
\ben
\cF(\{A\}) = \int d^4 x \;  \tr A_\mu(\bx) A_\mu(\bx) \; , 
\label{cF}
\een  
and the lattice analogue that we use is
\ben
\cF_L(\{U\})~=~\gb\, \real \Tr \sum_{\bm{x},n}c_n
\left[\sum_\mu f_\mu(U_\mu(\bx))^n\right]\,,
\label{gfn}
\een
where the $c_n$ are coefficients chosen so that terms in the
perturbative expansion of $\cF_L$ between $A^2$ and some higher power
are absent; and $f_{1,2,3} = 1/\chi$, $f_4 = \chi$. In practice, we
use
\ben
c_1 = 1,~c_2=-\frac{1}{8},~c_3 = \frac{1}{63},~c_4 = -\frac{1}{896} \; ,
\een
with $c_{n>4}=0$. This ensures that the definition of Landau gauge
agrees with that used in the perturbation theory up to and including
${\cal O}(g^8)$. This allows us to check the two loop perturbation theory for
$u_s,u_t$ and to deduce higher loop contributions using
the simulation. Our improved gauge-fixing
function is not the same as those described elsewhere
\cite{Lepage:1997}.
It does, however, correspond to the definition of Landau gauge used in
the perturbation theory, namely
\ben
\D^{(-)}_\mu A_\mu(\bx_\mu+\be_\mu/2) = 0,
\een
where $A_\mu(\bx+\be_\mu/2)$ is the perturbative gauge field. 

A Fourier accelerated algorithm is vital when fixing to Landau gauge,
which is otherwise prohibitively time consuming. To carry out the
Fourier transform on fields with twisted boundary condition we define
$\Ga(\bn,\bx)$ by
\ben
\Ga(\bn,\bx) = \Ga(\bn) \, \exp(2\pi i n_\mu x_\mu/N) \; ,
\een
where the constant matrices $\Ga(\bn)$ are defined in the Appendix.
For a twisted field $\bv(\bx)$ in the algebra of $SU(N)$ we define
\ben
v(\bn,\bx) = \Tr \left[\Ga^\dagger(\bn,\bx) \bv(\bx)\right] \; .
\label{fft0}
\een
Now $v(\bn,\bx)$ is periodic on the lattice and we can define the
periodic Fourier transform of $v(\bx)$, which can be implemented by
the FFT algorithm, by
\ben
\tv(\bn,\bm{k})= \sum_{\bm{x}} v(\bn,\bx) \exp(-2\pi i\bark_\mu x_\mu/N) 
\; ,
\label{fft1}
\een
where $k_\mu$ is given in terms of $\bark_\mu$ and $n_\mu$ in
Eqn.~(\ref{spectrum}) and hence the argument $\bm{n}$ on the LHS is
redundant and retained only as a reminder: $\bn$ can
be deduced from $\bm{k}$ but it is convenient to explicitly display
the dependence of $\tv$ on $\bn$. The inverse transform is then
\bea
v(\bn,\bx)&=&{1 \over V_{TW}}\sum_{\bar{\bm{k}}} \tv(\bn,\bm{k}) 
\exp(2\pi i\bark_\mu x_\mu/N)\;,\nn\\
\bv(\bx)&=&\sum_{\bm{n}} \Ga(\bn,\bx) v(\bn,\bx) \; ,
\label{fft2}
\eea
(see 
\cite{Drummond:2002yg}
for further details).

To maximise $\cF_L(\{U\})$ with respect to gauge transformations we
define the gauge velocity
\ben
\om(\bx)~=~T_a{\delta \over \delta \e_a(\bx)}\cF_L(\{U^{(g)}\})|_{\e=0}\;,
\een
and use steepest ascent. The gauge transformation $g(\bx)$ is given by
$g(\bx) = \exp \e_a(\bx)T_a$ and satisfies the twisted boundary
conditions
\ben
g(\bx+L_\mu\be_\mu)~=~\O_\mu\,g(\bx)\,\O^\dagger_\mu \; , 
\label{tgauge}
\een
and then
\ben
U^{(g)}_\mu(\bx)~=~g(\bx)\,U_\mu(\bx)\,g^\dagger(\bx+\be_\mu)\;.
\een
We, however, simulate with the periodic field $\{\hatU\}$ which
transforms under gauge transformations as
\ben
\hatU^{(g)}_\mu(\bx)~=~\hatg(\bx)\,\hatU_\mu(\bx)\,
\hatg^\dagger(\bx+\be_\mu)\;,
\een
with the rule
\ben
~\hatg(\bx)~=~g(\bx)\,,~~\hatg(\bx+L_\mu\be_\mu)~=~\hatg(\bx)\;.
\een
Note that, whilst the $\{\hatU\}$ transform effectively with a
periodic gauge transformation, the infinitesimal gauge transformation
still has the twisted spectrum derived from Eqn.~(\ref{tgauge}). The
outcome is that to maximise $\cF_L(\{U\})$ we apply successive
infinitesimal gauge transformations $\hatg(\bx) =
\exp(\e\,\hatom(\bx))$ to the configuration $\{\hatU\}$ where
\ben
\hatom(\bx) = \gb \sum_{n,\mu} nc_n f_\mu\,\cA\left[(U_\mu(\bx))^n - 
(U^\dagger_\mu(\bx-\be_\mu))^n\right] \; ,
\een
with $U_\mu(\bx) = \hatU_\mu(\bx)\hatO^\dagger_\mu(\bx)$ and $\e$ is a
small step size. The simulation uses the fields $\{\hatU\}$ and so an
explicit representation of the $\O_\mu$ is needed (such as in the
Appendix).

The Fourier acceleration of gauge fixing has been discussed in 
\cite{Davies:1988} 
and the Fourier accelerated gauge velocity is given by
\ben
\widetilde{\hatom}_{\mathrm{acc}}(\bn,\bm{k}) = 
\k(\bm{k})\,\widetilde{\hatom}(\bn,\bm{k}) \; ,
\een
where the tilde notation is defined in
Eqns.~(\ref{fft0}--\ref{fft2}). The acceleration kernel, $\k(\bm{k})$,
is given by
\ben
\k(\bm{k}) = \frac{1}{\bm{k}^2 + m^2} \; ,
\een
where $m$ is a mass parameter which may be set to zero for twisted
momenta as there is no zero mode. Note that since $\bm{k}^2$ depends
on $\bn$ for given periodic component $\barbbk$, then $\k(\bm{k})$
distinguishes between modes labelled by different $\bn$. This is
significant only for small $\bm{k}$ where, nevertheless, the
acceleration has the largest effect and careful control of the IR
behaviour of the algorithm is needed.

The gauge transformation $\hatg(\bx) =
\exp(\e\,\hatom_{\mathrm{acc}}(\bx))$ is constructed for small step
size $\e$. Because the exponentiation and Fourier acceleration are
expensive it is best to successively apply this gauge transformation
to the configuration $\{\hatU\}$ until a maximum of $\cF$ on the given
gauge orbit is achieved and then $\hatom_{\mathrm{acc}}$ is recomputed
and the procedure repeated. To speed up the process a sequence of
gauge transformations is constructed, $\hatg_m = \hatg^2_{m-1},~
\hatg_1 \equiv \hatg,~~m=1,\ldots,M$. The maximum on the given orbit
is determined using $\hatg_M$ and then successively refined using the
$\hatg_m$ in descending order in $m$. This reduces the number of
matrix multiplies required needed whilst allowing $\e$ to be chosen to
be very small. We chose $M=5$. The algorithm is applied until the
condition on the Fourier-accelerated velocity
\ben
\bb -{1 \over V}\sum_{\bm{x}} 
Tr[\hatom_{\mathrm{acc}}(\bx)^2] \eb^{1/2} \le \delta \; \,
\een
is satisfied. We found that $\delta \sim 10^{-4}$ to be more than
sufficient to fix the to Landau gauge very accurately.

We observe that (a) without Fourier acceleration convergence
is prohibitively slow, and (b) that convergence was improved by at
least a factor of 10, measured in computer time, when 4--twist
boundary conditions were used compared with the 2-- or 3--twist
boundary conditions. The reason for the latter result is unclear. It
is possible that tunnelling between different toron vacua is the most
strongly suppressed in the 4--twist case, but the improvement was
still much in evidence at large $\gb,~\gb \sim 30$, and from the mean
Polyakov line in the untwisted direction(s) there was no noticeable
signal that tunnelling was occurring in the 2-- and 3--twist
cases. The number of near zero modes is fewer in the 4--twist case but
there is no convincing argument that the effect should be so dramatic.

\subsection{Gribov copies}

Fixing $\{U\}$ uniquely to Landau gauge corresponds to finding the
global maximum of $\cF_L(\{U\})$. Other (local) maxima give rise
to lattice Gribov copies of this gauge. The global maxima lie in the
Fundamental Modular Region (FMR) of the Landau gauge cross-section and
since the FMR contains the identity we expect that our perturbative
calculations relate to fields gauge-fixed in this way. The mean link
values are closely related to $\cF_L$ at the global maximum and to use
Gribov copies to evaluate them will clearly give, on average, a lower
value for the tadpole coefficients than predicted by the perturbation
theory. We have not investigated the \textit{r\^ole} of Gribov copies in the
simulation and so, in principle, the measured tadpole coefficients
will have a negative systematic error. However, we believe that the
effect is negligible for large enough $\beta$ and when fully twisted
boundary conditions are employed. We defer further discussion of this
point until the next section.

\subsection{Simulations}

The simulations were carried out for a range of $\beta$ values from
$5.2 \le \beta \le 30$ for $\chi=2$ which easily encompasses the
physical region. Because it is more physically relevant we
concentrated on the improved Symanzik action. The simulations were run
on 64 processors on the Hitachi SR2201 computer at the Tokyo Computer
Centre on lattice size $8^4$, and on 48 processors on the SunFire F15K
of the Cambridge-Cranfield High Performance Computing Facility on an
$8^3 \times 16$ lattice. As mentioned above, we used the 4--twist
boundary conditions in these calculations.

Using the actions specified on the RHS of Eqn.~(\ref{sc_actions}) the
self-consistent tuning of $(u_s,u_t)$ is fast. In fact, it is
immediate for the Wilson action since there is no explicit reference
to $(u_s,u_t)$ once the rescaling in Eqn.~(\ref{sc_couplings}) has been
done.  For the Symanzik action there is a residual dependence through
the counter-term $\Delta S_{SI}$ on $u_s$ but starting at ${\cal O}(g^2)$. In
this case, in the simulation a value for $u_s$ is chosen and the mean
spatial link is measured and the next value of $u_s$ inferred which is
used as the starting value for a new simulation. This process needs to
be iterated only two or three times for accurate convergence to the
self-consistent values of $(u_s,u_t)$ to be determined. For more on
the efficient self-consistent tuning of tadpole improvement factors in
Monte Carlo simulations, see
\cite{Drummond:2003qu}
and the linear map techniques in
\cite{Alford:2000an}.
It is, of course, an aim of this study to remove the need for such
tuning.

\begin{figure}[b]
\bec
\includegraphics[height=60mm,clip]{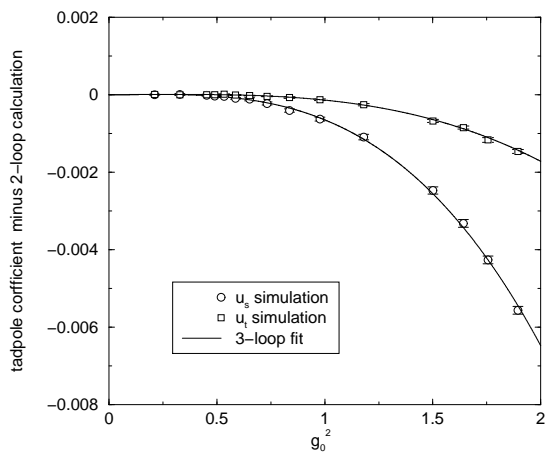}
\enc
\caption{\label{sim_l} Simulation of the Landau mean link tadpole
parameters for the Symanzik-improved action with $\chi=2$ versus
self-consistent coupling $g^2$ with second order perturbation theory
subtracted.  The solid lines show fits with three loop, ${\cal O}(g_0^6)$
parametrisation. For fit results see text.  }
\end{figure}

Precision comparison of perturbative and simulation mean links
requires consideration of finite size effects (FSEs), and we would
like to compare the two for a given volume. From looking at the
variation with $L$ of the perturbative results, the major finite size
effect was found to be from the one loop, ${\cal O}(g_0^2)$ contribution and
the simulation agreed accurately with the $L=8$ prediction. Already
for $L\ge 8$ the higher-order coefficients show very little
sensitivity to $L$ and so for comparing with simulation we set them at
their large-$L$ values. We then compare the results of the simulation
with the calculated two loop expansion. In Fig.~\ref{sim_l} we plot
\ben
u_- = u^{\mathrm(MC)} - \left( 
1 + b_l^{(1)} g_0^2 + b_l^{(2)} g_0^4 
\right) \; ,
\een
the measured mean link with the two loop perturbative prediction
subtracted, against $g_0^2$ for $\chi=2$.

The object is to show that the simulation is very accurate and agrees
very well with the perturbative prediction for small enough $g_0^2$
and to also infer the size of the deviation at larger $g_0^2$ due to
three loop and higher order terms, \textit{i.e.} at $g_0^6$ and
higher. A polynomial fit will give a good estimate for higher
coefficients in the self-consistent perturbation series.  Care must be
taken when fitting polynomials to data in a given range of the
expansion parameter.  If the true coefficients are such that in this
finite window there is an approximate cancellation between some
combination of terms, then {\em all} such terms must be included in
the fit function.  If not, as might be the case when trying
polynomials of increasing order, marked instability in goodness of fit
and even low order coefficients will be seen. Inclusion of all orders
can be approached using `Bayesian' techniques
\cite{lepage01}.
These did not, however, appear necessary in describing our data, and
the results quoted come from straight polynomial fits at the given
order.

Despite comparing simulations with the $L=8$ perturbation theory,
there can, however, still be slight FSEs as tunnelling has forced us
to use different boundary conditions in the simulation and
perturbation theory (which matters at finite $L$).  To allow for this
small discrepancy, we fit polynomials of the form
\ben
u_- = \delta b^{(2)}_l g_0^4 + b^{(3)}_l g_0^6 \; ,
\een
where $~\delta b^{(2)}_i$ is the finite size correction at two loops.
The quality of the fits is good down even to gauge couplings $\beta
\sim 5-6$, with no sign of terms of ${\cal O}(g_0^8)$. The
$\chi^2/\mathrm{d.o.f}$ are both $\sim 0.6$ and the fit coefficients
are
\begin{eqnarray}
\delta b^{(2)}_s & = & 0.00032(4) \; , ~~~~ b^{(3)}_s = -0.00097(3) \; ,
\nn \\
\delta b^{(2)}_t & = & 0.00015(1) \; , ~~~~ b^{(3)}_t = -0.00029(1) \; .
\end{eqnarray}
The finite size correction corresponds to a $2\%$ effect in the
two loop calculation. This is certainly very reasonable. The
measurement of the three loop coefficients is accurate to $3\%$ and we
should also expect a finite size error of a similar order.

\section{Discussion and conclusions}
\label{sec_conc}

Tadpole improvement via Landau mean link factors is becoming
increasingly important in lattice field theory, but there are
computational costs in tuning the mean-link factors to their
self--consistent values. In some cases, such as the small pure gauge
theory lattices used in this work, the costs are bearable. In other
cases they become significant. For example, highly anisotropic
lattices, such as would be used for finite temperature studies, would
need to be tuned at zero temperature for a physical spatial volume.
The very large time extent then required makes numerical tuning a
major study in itself.

In this paper we calculate the mean link in Landau gauge
perturbatively to two loop order. By comparing this to simulations, we
have shown that this result predicts the mean link in Landau gauge to
a high accuracy. Use of the formulae obtained remove the need for the
expensive tuning.

We have used the automatic generation of the perturbative vertices to
calculate the self-consistent tadpole-improvement parameters
$u_s$,~$u_t$ for the anisotropic Wilson and Symanzik-improved gluon
action to two loops as a function of the lattice coupling $g_0^2$ and
the anisotropy coefficient $\chi$. The tadpole-improved actions,
Eqn.~(\ref{actions}) are parametrised by $g_0$ and $\chi_0 =
\chi\,u_s/u_t$. The one and two loop coefficients are given in
Tables~\ref{tab_oneloop} and~\ref{tab_twoloops}, respectively, for
different $\chi$.  The calculation was done in Landau gauge with
twisted boundary conditions in the $\mu = 1,2$ directions so that
there was no zero mode. Whilst the one loop and figure-of-eight
integrals were done with mode summation on an $L^4$ lattice, the major
two loop numerical Feynman integrations were done using parallel {\sc
  Vegas}
\cite{lepage78,numrec} 
on an $L^2 \times \infty^2$ lattice. For $L>16$ (with appropriate
``squashing'' of the momentum variables
\cite{luwe})
there was no observable $L$-dependence. The coefficients of $g^2$ and
$g^4$ for the direct perturbation series and of $g_0^2$ and $g_0^4$
for the self-consistent perturbation series are given in
Tables~\ref{tab_oneloop}--\ref{tab_contribs} for different
anisotropies, $\chi$. Details of the component calculations are also
given in these tables to help facilitate the future calculation of
different but related quantities. Interpolating fits were obtained to
these coefficients as a function of $\chi$ (Table~\ref{tab_expansions}).

The simulation was done using a second order Langevin algorithm for
the field update and the Landau gauge is fixed for each configuration
using a Fourier accelerated steepest ascent method to maximise the
(improved) gauge function $\cF({\hatU})$. 

We found that fixing to Landau gauge was very slow unless
Fourier acceleration was used, and that by using twisted boundary
conditions in all four directions (4--twist) the convergence of the
algorithm was improved by at least a factor of ten compared with
applying the twist in just two or three directions (2-- or 3--twist).
The reason for this is not clear but may have a connection with the
suppression of tunnelling between different toron vacua, although no
clear signal supporting this surmise was found.  Practically, this
effect was welcome in the isotropic case, and essential for $\chi=2$
where the convergence was otherwise prohibitively slow.

We did not investigate the \textit{r\^ole} of lattice Gribov copies in fixing
to Landau gauge by maximizing $\cF_L(\{U\})$ in Eqn.~(\ref{cF}). There
has been a long discussion in the literature concerning the
\textit{r\^ole} of Gribov copies
\cite{mitr:1996,mitr:1997,boea:2000,bock:2001,olsi:2001,olsi:2003}.
In principle, the effect of selecting a local rather than the global
maximum of $\cF_L$ will be to reduce the measure values of the tadpole
improvement coefficients. We believe that the effect in this
calculation is negligible for a number of reasons. We do not expect a
problem with Gribov copies at large-$\beta$ but some effect may occur
as the physical region is approached
\cite{Hart:1997ar}.
However, the use of twisted boundary conditions is likely to suppress the
discrepancy due to the Gribov problem. This is because there is no
zero-mode with such boundary conditions and by using a twist in all
four directions the number of low-lying modes is reduced. In
\cite{boea:2000} 
the effect of the zero-mode was shown to be very important in the
$U(1)$ case and by performing the extra gauge transformation to remove
its effect, the ZML gauge, the distribution of mean link values
measured was greatly reduced. Another effect is due to double Dirac
sheets (DDS) which have zero action but are strongly correlated in the
$U(1)$ theory with a reduction in the mean link value and which are
mooted to also be relevant in $SU(N)$ theories. We see no evidence for
DDS contributions in that we see no negative spikes of the nature
clearly observed in the $U(1)$ case 
\cite{boea:2000}.
However, we have no quantitative analysis to support this
remark. 

If there are Gribov effects in our simulation then these will be
included in the spread of results as a systematic error and are
therefore to a great extent included in our quoted error which we
perforce analysed by purely statistical methods
\cite{Cucchieri:1997dx}.

The Landau gauge function is improved; it is chosen so that it corresponds
to the lattice Landau gauge used in the perturbative calculation up to
term of order $g_0^{10}$.  Thus, perturbation theory and the
simulation should agree up to and including four loops.

The simulations were done on an $L^3 \times T$ lattice with $L=8,
T=16$ for the Symanzik-improved action with $\chi=2$ whch corresponds
to an equal-side hypercubical lattice when measured in physical units.
The couplings used were in the range $5.2 \le \beta \le 30$. The
simulation boundary conditions were twisted in all four directions,
and in order to compare with calculation an adjustment for finite
$L,T$ effects is needed. Finite size effects were only appreciable in
the one loop coefficients. Accordingly, we used one loop coefficients
calculated on a finite lattice, and two loop data for infinite volume.
The higher order perturbative coefficients that are extracted may thus
change slightly for larger $L$.

We fitted the data (less the two loop perturbative result) with a
polynomial containing terms in $g^4$ and $g^6$. This described the data
extremely well over the full range of gauge couplings. The first term
in the fit function allowed for a small finite size correction to the
two loop perturbative result.  This finite size discrepancy
corresponds to a $2\%$ correction to the two loop calculation. This is
certainly very reasonable. The measurement of the three loop
coefficients is accurate to $3\%$ and we should also expect a
finite size error of a similar order. The main conclusion is that even
in the physical region $\beta \sim 5-6$ perturbation theory works very
well indeed for the Landau mean link and, moreover, there is no
observable deviation from a three loop, ${\cal O}(g_0^6)$ perturbative
approximation. This is very encouraging for the accurate design of QCD
actions and the corresponding perturbative analyses based upon them.

In principle, a full simulation analysis would give three loop numbers
for a range of $\chi$ values, enabling a fit of the same form as is
given in Eqn.~(\ref{ufit}). However, these coefficients are seen from
the figure to be small even in the physical region, and so we conclude
that $u_s$ and $u_t$ are approximated sufficiently accurately by the
two loop expressions for all practical purposes.

Finally, we remark that it could be possible to calculate $u_s$ and
$u_t$ by simulation using stochastic perturbation theory. However,
although it has proved very successful for the expansion of the mean
plaquette, the gauge fixing required here makes this approach
extremely slow. Nonetheless, in general, the double expansion of the
action and then the gauge potential itself as power series in the
coupling required for the stochastic evolution equations is a task to
which the {\sc Python} code can be readily adapted.

\begin{acknowledgments}

The authors would like to thank I.T. Drummond for useful discussions.
We acknowledge the work of A.J. Craig and N.A. Goodman. The
simulations were performed on the SunFire F15K computers of the
Cambridge-Cranfield High Performance Computing Facility and on the
Hitachi SR2201 of the University of Tokyo Computing Centre, and
the authors gratefully thank both facilities for help and provision of
resources. A.H. is supported by the Royal Society.

\end{acknowledgments}

\bibliographystyle{h-physrev4}
\bibliography{tadpole_refs}

\appendix

\section{Twisted boundary conditions}
\label{app_twbc}

We give here a brief description of the boundary conditions. Further
details can be found in
\cite{luwe,Drummond:2002yg}. 

For an orthogonal twist the twisted boundary condition for gauge
fields (and potentials) is
\ben
U_\mu(\bx+L_\nu \be_\nu) = \O_\nu U_\mu(\bx)\O^{-1}_\nu\;,
\label{tw}
\een
where the twist matrices $\O_\nu$ are constant $SU(N)$ matrices which
satisfy
\ben
\O_\mu\O_\nu = z_{\mu\nu}\O_\nu\O_\mu
\een
and $z_{\mu\nu}=\exp(2\pi i n_{\mu\nu}/N)$ is an element of the centre
of $SU(N)$. The particular boundary conditions imposed are uniquely
specified by the antisymmetric integer tensor $n_{\mu\nu}$,
$n_{\mu\nu} \in \cZ_N$
\cite{gonzalez97}. 
The twisted boundary conditions can be chosen to apply in two
directions only, taken here to be the (1,2) directions, but can also
be applied to the 3 and 4 directions if required. Our implementation
of these boundary conditions correspond to orthogonal twists since
$\kappa \equiv \e_{\mu\nu\s\ro} n_{\mu\nu}n_{\s\ro} = 0~\mod~N$, and
consequently only configurations with integral topological charge,
$Q$, will occur 
\cite{gonzalez97}; 
the perturbation theory is then correctly associated with the $Q=0$
sector.  In this case $\O_3,\O_4$ can be expressed in terms of
$\O_1,\O_2$ once the values of $n_{\mu \nu}$ are given. We use
\ben
\O_1=
\bb\ba{ccc}z&0&0\\0&1&0\\0&0&z^*\ea\eb\;,~~
\O_2=
\bb\ba{ccc}0&1&0\\0&0&1\\1&0&0\ea\eb\;,
\een
The different choices of boundary condition are then given by
assignments to $\O_3$ and $\O_4$ as follows:
\ben
\ba{l@{~~~~}l@{~~}c}
\O_3 & \O_4 & (n_{12},n_{13},n_{14},n_{23},n_{24},n_{34}) \\
\hline
\unit & \unit & (1,1,0,0,0,0)
\\
\O_1^\dagger\O_2^\dagger & \unit & (1,1,0,1,0,0)
\\
\O_1^\dagger\O_2^\dagger & \O_1\O_2^\dagger & (1,1,1,1,1,1)
\ea
\een
We refer to these choices as 2--,3--,4--twist boundary conditions,
respectively.

If the lattice has extent $L_\mu$ in the $\mu$--direction, the
momentum spectrum is
\begin{eqnarray}
k_\mu & = & \frac{2 \pi \bar{k}_\mu}{L_\mu} + \frac{2 \pi n_\mu}{NL_\mu}
\; , ~~~~  \bar{k}_\mu,n_\mu \in \mathbb{Z} \;,
\nonumber \\
&&  0 \le n_\mu < N \; , ~~~~ 0 \le \bar{k}_\mu < L_\mu \; .
\label{spectrum}
\end{eqnarray}
For 2--twist boundary conditions, $n_3=n_4=0$. The 3--twist boundary
conditions are given by $n_3 = -(n_1 + n_2)$, $n_4 = 0$ and the
4--twist ones by $n_3 = -(n_1 + n_2)$, $n_4 = n_1-n_2$. In all cases
we exclude the zero mode $n_1 = n_2 = 0$.

Negative momentum in these directions is $-k_\mu =
(-\bar{k}_\mu,-n_\mu)$, adding appropriate multiples of $L_\mu$ and
$N$ respectively to remain in the ranges defined above.

Twisted boundary conditions thus replace the $N^2-1$ colour components
of an $SU(N)$ field by a similar number of extra momentum components,
interstitial to the usual reciprocal lattice. The Fourier expansion of
a gauge potential is now
\begin{equation}
A_\mu(\bm{x}) = \frac{1}{V_{TW}} \sum_{\bm{k}}
e^{i \bm{k} \cdot \bm{x}} \Gamma(\bn) 
\tilde{A}_\mu(\bm{k}) \; ,
\label{eqn_tw_potl}
\end{equation}
where $V_{TW} = N \prod_\mu L_\mu$ is the twisted volume of the
lattice.  The sum over momentum indicates a sum over all $\bar{k}_\mu$
and the components of the twist vector $\bn \equiv (n_1,n_2)$.  The
$N^2-1$ $SU(N)$ twist matrices are given in terms of
$\O_{1,2}$ by
\ben
\Ga(\bn) = z^{(n_1+n_2)(n_1+n_2-1)/2}\O_1^{-n_2}\O_2^{n_1}\;,\label{Gan}
\een
where $z = \exp(2i\pi/N)$ is an element of the centre of $SU(N)$.

We need only the trace algebra associated with these.  Defining the
symmetric and antisymmetric products of twist vectors
\bea
(\bm{n},\bm{m}) & = & n_1 m_1 + n_2 m_2 + (n_1 + m_1)(n_2 + m_2) \; , 
\nn \\
\la \bm{n},\bm{m} \ra & = & n_1 m_2 - n_2 m_1 \; ,
\label{eqn_tw_prods}
\eea
we obtain
\begin{eqnarray}
\Ga(\bn)&=&\1~~~\bn = \b0~\mod~N \; ,
\nonumber \\
\mathop{\mathrm{Tr}} \left[ 
\Ga(\bn) \right] & = & 0~~~\bn \ne \b0~\mod~N \; ,
\nonumber \\
\Ga(\bn)^\dagger&=&z^{-\shalf(\sbn,\sbn)}\Ga(-\bn) \; ,
\nonumber \\
\Ga(\bn^\pr)\Ga(\bn)&=&z^{\shalf(\la \sbn^\pr,\sbn\ra-(\sbn^\pr,\sbn))}
\Ga(\bn^\pr+\bn) \; ,
\nonumber \\
\Rightarrow
\mathop{\mathrm{Tr}} \left[ \Gamma(\bn^\pr) \Gamma(\bn) \right]
& = & N z^{\frac{1}{2}(\sbn,\sbn)} \delta_{\sbn,-\sbn^\pr} \; ,
\label{eqn_tw_alg}
\end{eqnarray}
where $(\bn~\mod~N)$ is understood to apply to each component,
$n_{1,2}$, as is the $\delta$-function. The argument of $\Ga$ is
evaluated $\mod~N$.

Note that for clarity in the main text we sometimes replace the twist
vectors in expressions such as Eqns.~(\ref{eqn_tw_prods},\ref{eqn_tw_alg}) 
with their associated momenta. In each case, the argument is understood 
to be the twist vector alone.

Note also that in Eqn.~(\ref{eqn_tw_potl}) the location of the gauge
field, $\bm{x}$, is taken to be midway along the associated
link. The components, $x_\mu$ are often then expressed as integer
multiples of units of half a lattice spacing.

\end{document}